\documentclass[11pt]{article}
\pdfoutput=1
\newcommand{\Comment}[1]{{}}
\usepackage{amsfonts,amsthm,amsmath,amssymb,slashed}
\usepackage[textwidth = 430 pt, textheight = 630 pt]{geometry}
\usepackage{color}
\usepackage{booktabs}
\usepackage{subcaption}
\usepackage[export]{adjustbox}

\Comment{\usepackage{color}
\definecolor{MyDarkBlue}{rgb}{0.15,0.15,0.45}
\usepackage[linktocpage=true]{hyperref}
\hypersetup{
colorlinks=true,
citecolor=MyDarkBlue,
linkcolor=MyDarkBlue,
urlcolor=MyDarkBlue,
pdfauthor={Horatiu Nastase},
pdftitle={Reheating in holographic cosmology and connecting to MSSM constructions for particle physics, with a cosmological constant},
pdfsubject={hep-th}
}
\usepackage[utf8]{inputenc}
\usepackage[numbers,sort&compress]{natbib}
\usepackage{hypernat}}
\usepackage{graphicx}
\usepackage{cite}
\usepackage[colorlinks=true]{hyperref}
\hypersetup{
  allcolors = blue,
}

\newcommand\ignore[1]{}
\def\one{{\,\hbox{1\kern-.8mm l}}}

\def\Tr{{\rm Tr\, }}

\def\a{\alpha}\def\b{\beta}

\def\d{\partial}

\def\Tr{\mathop{\rm Tr}\nolimits}

\newcommand{\Cset}{{\,\,{{{^{_{\pmb{\mid}}}}\kern-.45em{\mathrm C}}}}}

\newcommand{\be}{\begin{equation}}
\newcommand{\bea}{\begin{eqnarray}}

\newcommand{\ee}{\end{equation}}
\newcommand{\eea}{\end{eqnarray}}

\providecommand{\lsim}{\lesssim}
\providecommand{\gsim}{\gtrsim}

\begin{document}

\renewcommand{\thefootnote}{\fnsymbol{footnote}}

\makeatletter
\@addtoreset{equation}{section}
\makeatother
\renewcommand{\theequation}{\thesection.\arabic{equation}}

\rightline{}
\rightline{}




\begin{center}
{\LARGE \bf{\sc Reheating in holographic cosmology and connecting to $\Lambda$-MSSM constructions for particle physics}}
\end{center}
 \vspace{1truecm}
\thispagestyle{empty} \centerline{
{\large \bf {\sc Horatiu Nastase${}^{a}$}}\footnote{E-mail address: \Comment{\href{mailto:horatiu.nastase@unesp.br}}{\tt horatiu.nastase@unesp.br}}
                                                        }

\vspace{.5cm}

\centerline{{\it ${}^a$Instituto de F\'{i}sica Te\'{o}rica, UNESP-Universidade Estadual Paulista}}
\centerline{{\it R. Dr. Bento T. Ferraz 271, Bl. II, Sao Paulo 01140-070, SP, Brazil}}

\vspace{1truecm}

\thispagestyle{empty}

\centerline{\sc Abstract}

\vspace{.4truecm}

\begin{center}
\begin{minipage}[c]{380pt}
{\noindent In this paper we model the transition from the non-geometric holographic cosmology regime, to the usual radiation dominated (RD) cosmology, 
known as reheating in the case of (perturbative) inflation. We find that we can easily transition into any MSSM construction of intersecting D6-branes, 
via $\a'$ corrections in the 3 dimensional field theory as well as in cosmology, followed by cosmological reheating via S-NS5-branes. Moreover, we can naturally
obtain the (true) cosmological constant $\Lambda$ of the observed order of magnitude. The resulting supersymmetry breaking is just outside the currently observed
energies.  The model is consistent with large (TeV scale) extra dimensions, but it prefers smaller ones, and the string scale is generically low. 
}
\end{minipage}
\end{center}

\vspace{.5cm}

\setcounter{page}{0}
\setcounter{tocdepth}{2}

\newpage

\renewcommand{\thefootnote}{\arabic{footnote}}
\setcounter{footnote}{0}

\linespread{1.1}
\parskip 4pt



\section{Introduction}

For the early Universe, one usually considers a "standard model" of cosmology, the $\Lambda$-CDM 
(cosmological constant plus cold dark matter), with inflation in the early stages.
It is usually stated that this is the only model consistent with all observations. However, that is not necessarily so. 
Based on the ideas of the AdS/CFT correspondence \cite{Maldacena:1997re}
(see the books \cite{Nastase:2015wjb,Ammon:2015wua} for a review), and on a holographic description of inflation 
defined by Maldacena in \cite{Maldacena:2002vr} via the "wave function of the Universe equals 
boundary partition function, $\Psi(\Phi)=Z(\Phi)$" proposal, a holographic cosmology paradigm 
was defined in \cite{McFadden:2009fg,McFadden:2010na}. In it, inflation (for perturbative gravity plus scalar)
is a particular corner, however a phenomenological approach was considered, for a non-geometric 
(strong gravity) phase that replaces inflation, and with a field theory dual that is perturbative and 
calculable.\footnote{See the book \cite{Nastase:2019mhe} for a review of the applications of string 
theory to cosmology, including the AdS/CFT correspondence and holographic cosmology.}

In \cite{Afshordi:2016dvb,Afshordi:2017ihr} it was shown that the predictions of the model for the 
CMBR fluctuations, which are parametrically different (different functions of the parameters)
than the ones of $\Lambda$-CDM plus inflation, nevertheless fit it as well as the inflationary ones 
within experimental errors (and improving the error bars would allow one to choose one over the other).
In particular, from the very detailed data analysis in \cite{Afshordi:2016dvb,Afshordi:2017ihr}, we note that
$\chi^2$ for the two models differs by 0.5 out of a total of 824, a fit that means they are indistinguishable 
as far as the data is concerned. Further, in \cite{Nastase:2019rsn} 
(with full details for the letter in \cite{HNSoon}), it was shown that the phenomenological 
non-geometric phase solves the usual puzzles of hot Big Bang solved by inflation, as well as the latter, 
albeit in somewhat different ways. In \cite{Nastase:2018cbf} it was argued that the cosmological 
constant problem is easily understood in the field theory dual of this phenomenological holographic 
cosmology, and in \cite{Bernardo:2018cow} an attempt was made to obtain the top-down holographic
model that would reproduce the CMBR results; while it did not reproduce them, it opened the door for other potential top-down constructions.
Thus the only remaining issue that distinguishes inflation from holographic cosmology in terms of 
experimental results is the transition to the usual, radiation dominated (RD) cosmology, which in 
inflation is called reheating. In this paper we address this issue. 

To do so however, we need to consider the coupling of gravity to a construction for the standard model, 
or more precisely its extension, so a $\Lambda$-MSSM construction (cosmological constant plus 
Minimal Supersymmetric Standard Model). Since we are in string theory, where holography is defined 
precisely (at the "top-down" level), this analog of reheating must be understood from coupling 
the construction for holographic cosmology to a (string theoretic) construction for $\Lambda$-MSSM. 
Thus we need to understand a (perhaps general) class of string constructions for particle physics
(as is usually the case in inflationary reheating), together with a cosmological constant, and the 
supersymmetry breaking it is associated with. Here we will need to deal with the apparently very difficult problems 
of the extremely low cosmological constant (the "cosmological constant problem") and the fact that 
supersymmetry is not yet observed at particle accelerators. Some people say supersymmetry is ruled out, but really, 
only "natural" constructions are ruled out, yet a "top-down" string construction could in principle 
look unnatural from the low energy point of view, so really, all we know for sure (completely model-independent) 
is that the energy scale of supersymmetry breaking is higher than the energy scale of current accelerators. 
We will see that our construction for reheating within a certain class of string theory
constructions for $\Lambda$-MSSM has some general lessons to teach us about both problems. 

The paper is organized as follows. In section 2, we review holographic cosmology, and find that we can modify the best fit to the CMBR
in \cite{Afshordi:2016dvb,Afshordi:2017ihr} to a best fit to a supersymmetric model. In section 3, we define the model for reheating, and 
transition to the usual RD cosmology, with a MSSM construction coming from intersecting D6 branes on a $CY_3$. We review the 
non-geometric phase in a calculable toy model with only $N$ D2-branes, and reheating happens through S-NS5-branes that create the 
intersecting D6-branes. In section 4, we describe the cosmology from the point of view of the 3 dimensional Euclidean field theory. 
In section 5, we first consider the relation of the cosmological constant to the 3 dimensional field theory, which then allows us to find the 
reheating temperature, and to find that we can obtain the experimentally observed value for the cosmological constant. Then we 
find that we naturally have a supersymmetry breaking at a scale of about 30 TeV. Finally, from the dynamics of the compact space, 
and the conditions to imposed by cosmology, we find constraints on the radius of the extra dimensions and the string scale. In section 6, 
we conclude.

\section{Review of holographic cosmology and CMBR fits}

In \cite{McFadden:2009fg,McFadden:2010na}, a phenomenological model of holographic cosmology was defined. 
Consider a cosmology that in a geometric phase would have metric and scalar given by (background plus fluctuations)
\bea
ds^2&=&-dt^2+a^2(t)[\delta_{ij}+h_{ij}(t,\vec{x})]dx^i dx^j\;,\cr
\Phi(t,\vec{x})&=&\phi(t)+\delta\phi(t,\vec{x})a\;.\label{cosmology}
\eea

One first performs a Wick rotation ("domain wall/cosmology correspondence"), $t=-iz$, together 
with the rotation $\bar \kappa^2=-\kappa^2\;,\;\;\; \bar q =-iq$ in gravity, 
inducing the rotation $\bar q=-iq\;,\;\;\;\; \bar N=-iN$ in field theory, where $\kappa $ is the 
gravitational Newton's constant, $q$ is a momentum scale and $N$ the rank of the gauge group of the field theory. 
A holographic calculation, either based on Maldacena's $\Psi(\Phi)=Z(\Phi)$ map, or an a 
priori one based on a Hamiltonian formalism \cite{Papadimitriou:2004ap,Papadimitriou:2004rz,Skenderis:2000in} 
gives the CMBR scalar and tensor power spectra,
\bea
 \Delta_S^2(q)&\equiv& \frac{q^3}{2\pi^3}\langle \zeta(q)\zeta(-q)\rangle \cr
 \Delta_T^2(q)&\equiv & \frac{q^3}{2\pi^3}\langle \gamma_{ij}(q)\gamma_{ij}(-q)\rangle\;,\label{deltaST}
\eea
as
 \bea
 \Delta_S^2(q)&=& -\frac{q^3}{16\pi^2{\rm Im}B(-iq)} \cr
 \Delta_T^2(q)&= &-\frac{2q^3}{\pi^2{\rm Im}
 A(-iq)}\;,\label{holodelta}
\eea
where $A$ and $B$ are 
the coefficients of the expansion of the energy-momentum tensor two-point function into 
2 given Lorentz structures (projectors), 
\be
\langle T_{ij}(\bar q)T_{kl}(-\bar q)\rangle =A(\bar q) \Pi_{ijkl}+B(\bar q)\pi_{ij}\pi_{kl}\;,\label{TAB}
\ee
with $\pi_{ij}$ and $\Pi_{ijkl}$ being dimensionless projectors. 

Note that holography is usually considered for asymptotically AdS geometries, but actually, applies more generally. 
In  \cite{Maldacena:2002vr} was considered the de Sitter case corresponding to inflation, and since \cite{Itzhaki:1998dd} 
the cases of nonconformal D-branes, leading to domain walls, was considered. This was eventually generalized to this
generic cosmological case.

In the case of inflation, the above cosmology is geometric (meaning, $h_{ij}$ and $\delta \phi$ 
are small fluctuations around the background, and gravity is weakly coupled), but the corresponding 
field theory dual to it would be strongly coupled, so not that interesting. 

The interesting case is the opposite, of strongly coupled gravity, or non-geometric phase, 
when strictly speaking the cosmology (\ref{cosmology}) is not valid (is meaningless), but rather we need 
to describe the model through its field theory dual, which becomes perturbative. 

For this 3 dimensional YM field theory dual, with Euclidean signature for the flat 3 dimensional space since its corresponding gravitational 
part had the same,
we take the most general construction consistent with the symmetries (gauge and generalized conformal), i.e., a phenomenological action, 
\bea
S_{\rm QFT}&=&\int d^3x \Tr\left[\frac{1}{2}F_{ij}F^{ij}+\delta_{M_1M_2}D_i\Phi^{M_1}\Phi^{M_2}+2\delta_{L_1L_2}\bar \psi^{L_1}\gamma^i D_i
\psi^{L_2}\right.\cr
&&\left.+\sqrt{2}g_{YM}\mu_{ML_1L_2}\Phi^M \bar \psi^{L_1}\psi^{L_2}+\frac{1}{6}g^2_{YM}\lambda_{M_1...M_4}\Phi^{M_1}...\Phi^{M_4}\right]\cr
&=&\frac{1}{g^2_{YM}}\int d^3x \Tr\left[\frac{1}{2}F_{ij}F^{ij}+\delta_{M_1M_2}D_i\Phi^{M_1}\Phi^{M_2}+2\delta_{L_1L_2}\bar \psi^{L_1}\gamma^i D_i
\psi^{L_2}\right.\cr
&&\left.+\sqrt{2}\mu_{ML_1L_2}\Phi^M \bar \psi^{L_1}\psi^{L_2}+\frac{1}{6}\lambda_{M_1...M_4}\Phi^{M_1}...\Phi^{M_4}\right]\;,\label{phenoaction}
\eea
plus a nonminimal coupling of gravity to the scalar $1/(2g^2_{YM})\int \xi_M R(\Phi^M)^2$,
where in the second line we have rescaled the fields such that we have a common $1/g^2_{YM}$ factor. As we see, the $SU(N)$ 
gauge theory has adjoint gauge field $A_i^a$, $N_s$ adjoint scalars $\Phi^{aM}$, $M=1,...,N_s$ and $N_f$ adjoint fermions 
$\psi^{aL}$, $L=1,...,N_f$, and has the property of "generalized conformal invariance", meaning in the second line we have the only dimensionful 
coupling $g^2_{YM}$, multiplying the action, whereas the Yukawa couplings $\mu_{ML_1L_2}$ and $\phi^4$ couplings $\lambda_{M_1...M_4}$
are dimensionless, and the fields have 4-dimensional dimensions. Therefore the action could be written as the dimensional reduction (on a circle)
of a conformal theory in 4 dimensions. 

The RG flow of the theory is dominated by the momentum dependence of the effective coupling 
\be
g^2_{\rm eff}=\frac{g^2_{YM}N}{q}.
\ee

In the non-geometric phase, with bulk cosmology defined through the perturbative gauge  theory  on the boundary, the 
holographic relations (\ref{holodelta}) lead to CMBR fluctuations power spectra which at 2 loops in perturbative YM theory
are \cite{McFadden:2009fg,McFadden:2010na}
\bea
\Delta^2_S(q)&=& \frac{\Delta_0^2}{1+\frac{gq_*}{q}\ln \left|\frac{q}{\b g q_*}\right|+{\cal O}\left(\frac{gq_*}{q}
\right)^2}\cr
\Delta^2_T(q)&=& \frac{\Delta_{0T}^2}{1+\frac{g_Tq_*}{q}\ln \left|\frac{q}{\b_T g q_*}\right|+{\cal O}
\left(\frac{g_Tq_*}{q}\right)^2}\;,\label{spectrumholo}
\eea
which are then fitted against observations. The fit is indistinguishable from the fit of $\Lambda$-CDM plus inflation, 
for which $\Delta^2_{S,T}(q)=\Delta_{0S,T}^2 q^{n_{S,T}(q)-1}$ ($\chi^2$ of 823.5 vs. 824.0) \cite{Afshordi:2016dvb,Afshordi:2017ihr}, 
and the fit constrains the parameters of the action (\ref{phenoaction}). We emphasize that {\em this is a different functional form 
than the one in $\Lambda$-CDM plus inflation, and there is no $n_s$ in it}. 

A point that sometimes generates confusion is that the $q^3$ in the definitions (\ref{deltaST}) and
(\ref{holodelta}) is cancelled in the spectrum (\ref{spectrumholo}). The reason is that the 3 dimensional energy-momentum tensor 
$T_{ij}$ has classical dimension 3, so $A(q)$ and $B(q)$ in (\ref{TAB}) start at $q^3$ in the perturbative loop expansion, with log corrections 
to the classical value, leading to (\ref{spectrumholo}).

Assuming no fermions, and that all Yukawas (and nonminimal 
couplings $\xi_M=\xi$) are equal ($=\lambda$) allows us to find the best fit values for $N, N_S, \xi,\lambda$ 
and $g^2_{YM}$ (or $g^2_{\rm eff}$ at $q=q_*$ corresponding 
to the horizon scale in the CMBR). 

From the best fit values for the parameters, one obtains that (for $\xi=0.133$ corresponding to $r=0.12$, and for $\lambda=1$)
$N\sim 3\times 10^3$, $N_S\sim 2\times 10^4$, and that $g^2_{\rm eff}\sim 1$
(more precisely, $f_1g^2_{\rm eff}\ln g^2_{\rm eff}$ (the parameter appearing in the expansion at two loops)$\sim 1$ at about $l=35$
($l=1$ corresponds to $q=q_*$), so all CMBR modes with $l<35$ (i.e., $q<q_{l=35}$) correspond to non-perturbative YM, and therefore 
are excluded from the fit to the 2-loop expressions (\ref{spectrumholo}). We note that $\xi=0$ is incompatible with the best fit, in this model
(more details for this, and for the fit can be found in \cite{Afshordi:2017ihr}).

Since time evolution in cosmology corresponds to inverse RG flow in field theory (see for instance \cite{Nastase:2019rsn,HNSoon}), 
meaning $q$ increase and, as in the case of (geometric) inflation, the first modes to leave the horizon during the nongeometric 
phase are the last ones to come back in during our times (so correspond to the horizon scale), we get the picture above: the low momentum 
modes $q<q_{l=35}$ are the large scale modes in the sky with $l<35$ and are non-perturbative. 

But that means that as $q$ increases in field theory (along the cosmological time evolution), 
$g^2_{\rm eff}$ becomes lower and lower, thus the gauge theory becomes more and 
more perturbative. By the gauge/gravity duality principle (i.e., assuming such a gauge/gravity 
duality construction is well defined) that means that the gravity theory is more and more non-perturbative. 

This certainly runs counter to the usual picture, in which gravity is stronger at the beginning, and then becomes weaker and 
weaker. That means that we need to reverse this flow: the analog of reheating, ending the non-geometrical phase and connecting 
to the usual RD cosmology, must reverse this flow, and take us back towards perturbative gravity (weaker and weaker). 

Here we have reviewed only the power spectra, in particular the scalar power spectrum that is already observed, but in holographic 
cosmology other observables, like non-Gaussianity have been calculated
(non-Gaussianity of exactly factorizable equilateral shape 
with $f_{NL}^{\rm equil}=5/36$ was obtained in \cite{McFadden:2010vh}), while also the results of tensor fluctuations above are 
different than in inflation. The prediction for for the tensor to scalar ratio $r$ will be shortly addressed in the next subsection. 
Future observations will be able to discern between the inflation and the above phenomenological model of 
holographic cosmology (though the larger holographic cosmology paradigm contains other models, including inflation). 

Together with the explanation of the usual puzzles of Hot Big Bang cosmology in \cite{Nastase:2019rsn,HNSoon}, and the reheating in this paper, 
holographic cosmology is on the same footing as inflation in explaining all the available data.

\subsection{Best fit for supersymmetric model}

Before we see how to do that, we make a small interlude, and note that the fit in \cite{Afshordi:2016dvb,Afshordi:2017ihr}, which 
is really a fit on the parameters in the functional form for  $\Delta_S^2(q)$, $\Delta_T^2(q)$, could be used to find parameters for a 
supersymmetric model, specifically with ${\cal N}=1$ supersymmetry, as needed for particle phenomenology, with a few modifications. 
Of course, the 3 dimensional field theory describes 4 dimensional (super)gravity only, and needs to be coupled to 
4 dimensional supersymmetric particle physics, 
but if we want ${\cal N}=1$ supersymmetry for the latter, we also must have it for the former, so we must see if the CMBR fit is 
consistent with ${\cal N}=1$ supersymmetry for the 3 dimensional field theory. 

If we have ${\cal N}=1$ supersymmetry, then by equating the number of on-shell bosonic degrees of freedom with the 
number of fermionic degrees
of freedom,
\be
N_f=N_s+1.
\ee

In \cite{Afshordi:2016dvb}, the matching was done with $N_f=0$ and $\mu=0$ (no Yukawas, only equal quartic 
coupling, and that with $\lambda=1$), but scalars have more pull than fermions in observations (as also noted there), 
so in principle we can still achieve matching to the experimental data.

The theoretical prediction for the tensor to scalar ratio is 
\be
r=32\frac{1+\sum_{M=1}^{N_s}(1-8\xi_M)^2}{1+2N_f+N_s}.
\ee

In particular, at large $N_s,N_f$, we have
\be
r\simeq 32 \frac{N_s}{2N_f+N_s}(1-8\xi)^2.
\ee

If one actually takes $N_f=0$, as considered in \cite{Afshordi:2016dvb}, the result for $r$ is also independent of $N_s$, and only fixes $\xi$ via 
$r\simeq 32 (1-8\xi)^2$. Of course, we have not seen tensors yet, we only have a bound for them. But assuming $r=0.12$ (corresponding 
to the experimental limit), we fix $\xi=0.133$.

If on the other hand we take 
$N_f\simeq N_s$ (since $N_s\sim 10^4$ from the other fit, in $N_f=N_s+1$ we can ignore the 1), we have 
\be
r\simeq \frac{32}{3}(1-8\xi)^2\;, 
\ee
so just replace $(1-8\xi)^2$ with $(1-8\xi^2)/3$ in the experimental fit to $r$. That means that again $\xi$ is fixed from $r$. We are left with the 
parameters $N,N_s,g^2_{YM}$ (and $N_f=N_s+1\simeq N_s$).

Given a (common) value for the $\phi^4$ coupling $\lambda$ and for $\xi$ (from the above constraint on $r$), 
the results for $N,N_s, g^2_{YM}$ are fixed from $\Delta^2_0,\ln \b, gq_*$; the last (third) of both will be ignored. 
At large $N,N_s,N_f$, from $\Delta_0^2=1/(4\pi^2N^2f_0)$, we have \cite{Afshordi:2016dvb} ($\Delta_0^2=64/(4\pi^2 N^2 {\cal N}_{(B)})$,
${\cal N}_{(B)}=1+N_s(1-8\xi)^2$)
\be
\Delta_0^2=\frac{64}{4\pi^2 N^2 N_s (1-8\xi)^2}\;,
\ee
so in the matching to experimental data for $\Delta_0^2$, we replace $N^2N_s\rightarrow 3N^2 N_s$ as being equal to the same 
number (since we replaced $(1-8\xi^2)$ to $(1-8\xi^2)/3$ in being equal to the same number).

On the other hand, we find that, for the best fit and at large $N, N_s$
($N_s\simeq N_f$), and {\em if we can still put the $\phi^4$ coupling $\lambda$
to 1, while we can neglect the Yukawa coupling $\mu$ as $\mu^2N_s^2\ll 1$ (we will shortly see when and how could this is possible)},
we can approximate the formulas in \cite{Afshordi:2016dvb}\footnote{The general formulas for a constant $\xi_M=\xi$ are:
\bea
f_1&=&-\frac{4}{3\pi^2{\cal N}_{(B)}}\left(N_f-2-2N_s+\frac{\sum_M\mu^2_{MM}}{2}-48\Sigma_\phi\right)\cr
a_0&=&-\frac{1}{24\pi^2{\cal N}_{(B)}}\left[16+3\pi^2-56 N_f-4\sum_M \mu^2_{MM}+3(8\xi-1)[8(\pi^2-6)\xi N_s -3\pi^2+112
+2\sum_M \mu^2_{MM}]\right.\cr
&&\left. +\pi^2(1-8\xi)^2\sum_{M_1M_2}\lambda_{M_1M_2M_3M_4}\right]\cr
\Sigma_\phi&=&\xi^2\left(2N_s+\frac{1}{2}\sum_M \mu^2_{MM}\right)\cr
\mu^2_{M_1M_2}&=& \sum_{L_1,L_2}\mu_{M_1L_1L_2}\mu_{M_1L_2L_1}.
\eea}
input into 
\be
\ln \b = \ln \frac{1}{|g|}-\frac{a_0}{f_1}-\frac{64\Sigma_\phi}{\pi^2 f_1{\cal N}_{(B)}}\ln \frac{Nf_1}{g}\label{lnbeta}
\ee
as
\be
\ln \b\simeq -\frac{3\pi^2}{96}\frac{N_s(1-8\xi)^2}{3-96\xi^2}+\ln N.
\ee
Note that we have neglected on the right-hand side terms of order 1, only for the purposes of analyzing the (leading) 
$N, N_s$ dependence; 
otherwise, from the fit to experimental data (see table I of \cite{Afshordi:2016dvb}), $\ln \b\simeq 1.014$, and 
$\ln 1/|g|\simeq \ln 1/0.01305\simeq 4.33$, so should definitely not be neglected. 

Both terms on the right-hand side are 1/3 of the terms in the purely bosonic formula obtained for the case in \cite{Afshordi:2016dvb}.
Since $1-8\xi\simeq 0$, we have $\xi\simeq 1/8$, and the change is not big, in particular $3-96\xi^2\simeq 
3/2$ and is unchanged. Then we can keep $N$ constant for the best fit value, in which case $N_s(1-8\xi)^2$
is unchanged, for both $\Delta_0^2$ and $\ln \b$, but both terms are changed, and the best fit value for $N_s$ is divided by 3 with respect 
to the $N_s=0$ case. 

That means for the supersymmetric case, we would have the best fit values 
\be
N\simeq 3000\;, \;\; N_f\simeq N_s\simeq 9 000.
\ee

We now comment on the possibility of still using a common $\lambda=1$, 
and neglecting the Yukawas $\mu_{ML_1L_2}$ in the 
general formulas.\footnote{I thank the anonymous referee for pointing out that this issue needs explaining.}
Since in a supersymmetric theory, both Yukawas and the $\phi^4$ coupling come from the same superpotential term, 
of the type
\be
W=\frac{\tilde \mu_{M_1M_2M_3}}{3}\Phi_{M_1}^a\Phi_{M_2}^b\Phi_{M_3}^c k_{abc}\;,
\ee
with $k_{abc}$ some adjoint coefficients, which leads to Yukawas $\tilde \mu_{M_1M_2M_3}\phi^a_{M_3}\psi^b_{M_1}\psi^c_{M_2}
k_{abc}$, and to an F-term potential $\sum_{M_1,M_2,M_3}
|\tilde \mu_{M_1M_2M_3}\phi_{M_2}^b\phi_{M_3}^c k_{abc}|^2$, so apparently to 
$\mu_{M_1M_2M_3}=\tilde \mu_{M_1M_2M_3}$ and a coefficient 
\be
\lambda_{M_1M_2M_3M_4}=\sum_M \tilde \mu_{MM_1M_3}\tilde \mu _{MM_2M_4}\;,
\ee
and it would seem that a constant $\tilde \mu$ leads to a constant $\mu$, and a constant $\lambda=N_s\mu^2$, so 
in $a_0$ both would contribute $\mu^2N_s^3$, except $\lambda$ is multiplied by $(1-8\xi)^2\simeq 0$, so could be 
neglected with respect to the first. In that case, 
we would obtain $a_0/f_1<0$ and $\Sigma_\phi/(f_1{\cal N}_{(B)})<0$ (and the argument of the $\ln$ is $>1$), such that 
the right-hand side of (\ref{lnbeta}) has 3 positive terms, which would be a contradiction, since $\ln 1/|g|>\ln |b|$, as we saw.

To see how this could be avoided, we note that in \cite{Afshordi:2016dvb} the $SU(N)$ adjoint indices have been omitted, 
however the Yukawas are multiplied by $\Tr[T_a T_b T_c]=(f_{abc}+d_{abc})/2$, therefore standing in for $k_{abc}$. 
In ${\cal N}=4$ SYM, the superpotential has instead the structure $\Tr[T_a [T_b,T_c]]=f_{abc}$, which in our case would 
imply antisymmetry of $\mu_{M_1M_2M_3}$ in $M_2M_3$, but for the ${\cal N}=1$ theory considered here, we could have the 
structure $\Tr[T_aT_bT_c]$, which is neither antisymmetric like $f_{abc}$, nor symmetric like $d_{abc}$. Then $\mu_{M_1M_2M_3}$ 
could be made such that it equals zero for $M_2\geq M_3$. In that case, 
$\sum_M\mu^2_{MM}=\sum_M \sum_{L_1,L_2}\mu_{M L_1L_2}\mu_{M L_2L_1}=0$, so the Yukawas would not contribute to $\ln \b$, 
as we wanted. Unfortunately\footnote{I thank the anonymous referee for pointing this out to us, in the original reference 
\cite{Afshordi:2016dvb} the condition was not explained.}, in \cite{Coriano:2020zap}, which gives the details of the
calculation in \cite{Afshordi:2016dvb}, it was explained that the calculation was actually done with antisymmetric Yukawas and 
totally symmetric $\lambda$'s, which excludes the case presented here (in the first version of the paper I was not aware of this 
very recent reference). One would need to do the calculation without this condition. However, I expect that not much will change
(perhaps even the same formulas holding, if the symmetry properties were not essential to the calculation of the 2-point functions), 
since, as I mentioned, the contribution of the fermions is smaller than the one for the scalars, so qualitatively we expect the 
fit to be able to accommodate supersymmetric theories. 

Continuing for the moment under the assumption that the same formulas hold in the case on coefficients without symmetry properties,
in order to have the contribution of the $\lambda$ term be $N_s^2$, as we assumed before, we need to have 
also $\tilde \mu_{ML_1L_2}\sim {\cal O}(1/\sqrt{N_s})$, which for the best fit value of $N_s\sim 9 000$ is about $1/100$, not too
fine-tuned. If nevertheless we want $\tilde \mu\sim {\cal O}(1)$, it would still work, except the fit would be modified: we would obtain, 
instead of (\ref{lnbeta}), $\ln \b \simeq-\frac{\pi^2}{48}N_s^2(1-8\xi)^2+\ln N+{\cal O}(1)$, so we would need to change the value of $\xi$
to make it much closer to 1/8 in order to get the same fit for $N,N_s$. 
As a final note, we have also the gauge supermultiplet, leading to scalar D-terms and Yukawas involving the gaugino, but those 
can be shown to modify only slightly the fit, since they are very sparse, so will be ignored.

We conclude that there is very likely to be  a supersymmetric model, with one ${\cal N}=1$ vector multiplet (one gauge field and one 
fermion), and $N_s\simeq 9 000$ (or in any case a very large number of) 
chiral multiplets (with one scalar and one fermion) that is consistent with the best fit data to the CMBR
from \cite{Afshordi:2016dvb}. The actual numbers from the fit will not be much relied upon, and will be only used as examples towards the 
end of the paper.

\section{Reheating and transition to RD cosmology}

We come back to the issue of the "reheating", i.e., the transition from the non-geometrical phase (strong gravity) to the RD cosmology. 
We call it reheating in that energy flows from the gravity+scalar modes into MSSM modes, and because of the 
analogy with inflation: indeed, in a simple toy model, we will see that we obtain $a(t)\propto t^7$, that would be a power-law inflation
in a geometric (weak gravity) phase.
To understand this transition, we need to embed the best fit for the phenomenological model into a class of "top-down" models. 
Otherwise, it is hard to see how one could reverse the RG flow direction in a purely phenomenological theory. 

Since we have seen that we can have a best fit supersymmetric model, we will start with the simplest model for holographic 
cosmology, the system of $N$ D2-branes, whose holography was first discussed in \cite{Itzhaki:1998dd}. 
In the field theory limit, on the 3 dimensional worldvolume of the D2-branes there is a vector multiplet and 3 chiral multiplets. 
From the best fit to CMBR, we see that we need to supplement it with a very large number of chiral multiplets. 

Let us assume that, instead of the usual $S^6$ transverse to (surrounding) the D2-branes, we have a generic Calabi-Yau space $CY_3$, 
thus preserving ${\cal N}=1$ supersymmetry. We will analyze the implications of this change shortly. 
Then, in order to preserve ${\cal N}=1$ supersymmetry and generate 3 dimensional chiral multiplets, we can wrap D6-branes $D6_i$ on surfaces 
$CY_{2,i}\subset CY_3$, {\em while leaving the radial direction (the one Wick rotated to time in cosmology) not wrapped} (so wrapping the 
3 directions of field theory, and of the D2-branes, and the 4 dimensions of the $CY_{2,i}$). 
The geometry on which we wrap is determined by the $N$ D2-branes, so the fields on $D6_i$ are still in the 
adjoint of $SU(N)$ (another way of saying is that for each wrapped $D6_i$ brane we obtain $N$ {\em fractional} (effective) D2-branes).
Note that therefore the solution with backreaction for the $D6_i$ branes 
will not related to the D2-D6 solution of Cherkis and Hashimoto \cite{Cherkis:2002ir}, since that
wraps a radial direction (though not the full transverse radial direction of the D2-branes). 

Now, the $S^6$ has positive curvature (so the Ricci tensor $R_{ij}\neq 0$), and in the $N$ D2-brane  solution there is flux going through 
the $S^6$. In the D2-brane solution, the flux in $(012r)$ goes away and dilutes to $r\rightarrow \infty$, 
and the Einstein equation on $S^6$ is $R_{ij}\propto T_{ij}(flux)$. But $CY_3$ is Ricci flat (so $R_{ij}=0$), so we must arrange for the flux 
of the wrapped $D6_i$ to cancel the D2-brane flux, both in (the Einstein equation on) $CY_3$ and in the 
$(012r)$ directions (we can let some flux leak in the $(012r)$ 
directions as $r\rightarrow \infty$, see later). This necessarily will be a complicated solution, so we will not attempt it. 

A note on notation: I spoke of D2-branes and $D6_i$ branes but, after the Wick rotation, or domain wall-cosmology correspondence, these 
are now Euclidean branes, since their worldvolumes do not encompass the cosmological time $t$. Nevertheless, I will not call them S-branes, 
since they are located at $r=0$, corresponding to $t=0$; I will reserve the term S-branes for branes located at an $r_0\neq 0$, corresponding 
to $t=t_0$.

\subsection{Toy model}

Of course, since the chiral multiplets far outnumber the multiplets coming from the $N$ D2-branes, their 
contribution is crucial to holography, and that will depend on the details of $D6_i$ branes, but as a simple toy model, let us consider 
just the $N$ D2-branes, and review what we know. 

From the paper \cite{Itzhaki:1998dd}, 
in the validity {\em of the limit} $r\rightarrow 0,\a'\rightarrow 0$ (with $U=r/\a'$ fixed and $g^2_{YM}=g_s/\sqrt{\a'}$),
the phase structure is as follows (in the paper, the phase structure is described in terms of 2 variables, $g^2_{\rm eff}$ and $N$). 
In terms of increasing $\log g^2_{\rm eff}$, or {\em decreasing} momentum scale $U$, and at 
large $N$: we have first perturbative $SU(N)$ SYM on the $N$ D2-branes, then the IIA D2-brane 
supergravity solution with no 1 in the harmonic function $H_2$ (for weaker SYM, stronger gravity), 
then the M2-brane $AdS_4$ supergravity (at even stronger gravity, the IIA string theory turns into M-theory), finally
with the SCFT on the $N$ M2-branes in the deep IR (at the lowest values of $U$, or largest values of $g^2_{\rm eff}$; this would correspond
to what we now call the ABJM model \cite{Aharony:2008ug}). 

The D2-brane supergravity solution with no 1 in the harmonic function $H_2$\footnote{A $p$-brane solution in string frame is generically 
$ds^2=H_p^{-1/2}dx_{||}^2+H_p^{+1/2}dx_{\perp}^2$, where the harmonic function $H_p=1+\tilde d_pq_p(\sqrt{\a'}/r)^{7-p}$, with 
$\tilde d_p$ some 
numbers and $q_p$ quantized charges. The decoupling limit, leading to gravity duals, corresponds usually to the case when the 1 in $H_p$ 
can be neglected.}
 (what we would call the "gravity dual") has domain of validity \cite{Itzhaki:1998dd}
\be
g^2_{YM}N^{1/5}\ll U\ll g^2_{YM}N.\label{range}
\ee
Note that $U$ is what we have called here $q$, namely the momentum scale of the field theory. For $U\gg g^2_{YM}N$, we obtain a 
perturbative SYM phase, corresponding in gravity to a non-geometric (strong gravity) phase.

However, note a strange thing: if now we consider $U$ so high, that it goes {\em outside the limit} $U=r/\a'$ fixed, namely 
$r\sim \sqrt{\a'}$ or higher 
(outside the limit when $r=(U\sqrt{\a'})\sqrt{\a'}$, but $U\sqrt{\a'}\rightarrow 0$), then we go back to the supergravity description in terms of 
harmonic function $H_2=1+$ corrections, so we are away from the $N$ D2-branes, where the space would be flat for $CY_3\rightarrow S^6$. 
So now we have {\em two} supergravity regimes sandwiching a SYM regime in terms of momentum scale $U$ or $q$. 

To summarize the $N$ D2-brane toy model so far: in terms of increasing $U$ or $q$, corresponding in our case to increasing time in cosmology,
after the ABJM model and the M2-brane $AdS_4$ supergravity ("gravity dual of 
ABJM"), we have the type IIA "gravity dual" (a geometric phase in cosmology), followed by perturbative SYM (a non-geometric phase in 
cosmology), but if we continue on, we will reach another phase, where we have gravity again, but with the harmonic function where the 1 
dominates over the other term, but where crucially now the field theory is {\em not decoupled} from gravity. 
Note that for a D2-brane ($p=2$), the harmonic function is 
\be
H_p=1+d_p g^2_{YM}N/(U^{7-p}\a'^2)
=1+d_2 g^2_{YM}N/(\a'^2U^5).
\ee

Then we see one possibility to reverse the RG flow and go to a geometric phase again: we need to consider the $\a'$ corrections 
for the field theory on the $N$ D2-branes. When we reach 
\be
U\equiv q=\frac{1}{\sqrt{\a'}}\;,
\ee
(remember that $U=r/\a'$, so this corresponds to $r\sim \sqrt{\a'}$)
the RG transformation law of the field theory changes. The effective self-coupling of gravity modes $g^2_{\rm eff}
=g^2_{YM}N/q$ is negligible, but now another effective coupling appears: string D-brane $\a' F_{\mu\nu}^3$ corrections
appear (from integrating out string modes on the D-brane), which means that corrections in this new dimensionless coupling,
\be
\tilde \lambda_{\rm eff}=(\a' q^2)\;,\label{lambdatil}
\ee
become important, though now the field theory on the D2-branes and 4-dimensional gravity are coupled. As we see, we have reversed the 
RG flow, meaning that the new effective coupling of the field theory, $\tilde \lambda_{\rm eff}$, increases as $q$ increases, as we wanted. 
Note that $g^2_{\rm eff}$ still continues to decrease, but is now nearly zero, so becomes irrelevant to the field theory dynamics.

However, in order to have a good gravitational description, as we want to transition to a standard cosmology, 
we need that the string coupling in 10 dimensions becomes weak, as well as 
having 10 dimensional curvatures be small in string units. Of course, in the case of $CY_3\rightarrow S^6$, at large enough $U=q$, or 
equivalently large enough $r=U\a'$, $H_2\simeq 1$, and the gravitational space is flat, meaning that we have again 10 dimensional 
small curvatures. 

We need to understand how much we need to flow in $q=U$ between the non-geometric phase and the new geometric phase. 
Since we have
\be
g^2_{YM}=\frac{g_s}{\sqrt{\a'}}\;,
\ee
when we reach $U=U_{\a'}=1/\sqrt{\a'}$, we reach $g^2_{\rm eff, \a'}=g^2_{YM}N/U_{\a'}=g_sN$, which is the {\em 4 dimensional} 
't Hooft coupling divided by $4\pi$ ($\lambda_{tH}$), 
{\em so it is a very small value, but not that small} (we need the 3 dimensional gauge theory to be in the deeply 
perturbative domain already at the beginning of the SYM
flow, when $U\ll 1/\sqrt{\a'}$ - since $U$ is fixed, but $\a'\rightarrow 0$). We must then flow up to $U\gg 1/\sqrt{\a'}$ by at least that much. 
We have then 
\be
g_sN=\frac{g^2_{eff, initial}}{U_{\a'}/U_{in}}\sim \frac{1}{U_{\a'}/U_{in}},\label{UUin}
\ee
which is the amount of RG flow from strong effective SYM coupling to the string scale.

\subsection{Review: pre-geometric phase}

Before continuing, we review the construction of the cosmology corresponding to the holographic dual of $N$ D2-branes, since it will 
be useful in the following. In our toy model, it would correspond to a geometric phase that precedes the non-geometric phase that was fitted 
against the CMBR. See also \cite{Nastase:2018cbf}. 

The $N$ D2-brane solution has dilaton and 10 dimensional string frame metric (for $\tilde p=2$)
\bea
e^{\phi}&\sim & \left(\frac{g^2_{YM}N}{U}\right)^{5/4}\frac{1}{N}\cr
\frac{ds^2_{10d,s}}{\a'}&\simeq &U^2\sqrt{\frac{U^{3-\tilde p}}{g^2_{YM}N d_{\tilde p}}}dx_{||}^2
+\sqrt{\frac{g^2_{YM}Nd_{\tilde p}}{U^{3-\tilde p}}}
\frac{dU^2}{U^2}+\sqrt{\frac{g^2_{YM}N d_{\tilde p}}{U^{3-\tilde p}}}d\Omega_{8-\tilde p}^2\cr
&=& \frac{U^2}{R^2}dx_{||}^2+R^2\frac{dU^2}{U^2}+R^2d\Omega_{8-\tilde p}^2\;,\label{ND2}
\eea
where we have defined $R^2=\a'\sqrt{d_{\tilde p} g^2_{YM} N/U}$. Then the 10 dimensional Einstein frame metric is 
\be
ds^2_{10d,E}=e^{-\phi/2}ds^2_{10d,s}=\lambda_{\rm eff}^{-5/8} \sqrt{N}ds^2_{10d,s}
=\lambda_{\rm eff}^{-5/8}\sqrt{N}\left(\frac{U^2}{R^2}dx_{||}^2+R^2\frac{dU^2}{U^2}+R^2
d\Omega_6^2\right)\;,
\ee
where we have renoted $\lambda_{\rm eff}=g^2_{\rm eff}$. 

That means that the radius of $S^6$ {\em in the 10d Einstein metric} is 
\be
R'=R\lambda^{-5/16}_{\rm eff}N^{1/4}\sim \sqrt{\a'} \lambda_{\rm eff}^{-1/16}N^{1/4}.\label{Rpr}
\ee

For KK compactification from $D$ dimensions to $d$ dimensions,
\be
\tilde g_{\mu\nu}^{(E,d)}=\Delta^{\frac{2}{d-2}}g_{\mu\nu}^{(E,D)}.
\ee

In our case, we want to compactify on the $S^6$, from 10 dimensions to 4 dimensions, so 
$\Delta^{\frac{2}{d-2}}=\Delta=R'^6$, we find
\bea
ds^2_{E,4d}&=& \Delta ds^2_{E,10d}=\lambda_{\rm eff}^{-\frac{6}{16}}N^{\frac{6}{4}}\lambda_{\rm eff}
^{-5/8}\sqrt{N}\left[\frac{U^2}{\lambda_{\rm eff}^{1/2}}dx_{||}^2+\lambda_{\rm eff}^{1/2}
\frac{dU^2}{U^2}\right]\cr
&=& N^2\left[U^2\lambda_{\rm eff}^{-3/2}dx_{||}^2+\lambda_{\rm eff}^{-1/2}\frac{dU^2}{U^2}
\right]\cr
&=& a^2(t)dx_{||}^2+dt^2\sim N^2[U^{7/2}dx_{||}^2+U^{-3/2}dU^2].
\eea

Identifying the resulting metric with the cosmological ansatz, we obtain
\bea
&& dt^2\sim U^{-3/2}dU^2\sim [d(U^{1/4}]^2\Rightarrow t\sim U^{1/4}\cr
&& a^2(t)\sim U^{7/2}\Rightarrow a(t)\sim U^{7/4}\sim t^7.
\eea
We see that this phase is a kind of power law inflation, however this corresponds to the geometric phase before the relevant non-geometric phase.

Note that the 4 dimensional gravity becomes weak at large $U$, meaning at large cosmological time $t$.
But note that the D2-brane background duality is such that, as usual, the 
{\em 10 dimensional string metric} has strong curvature (since the radius of the metric is still $R\propto \lambda_{\rm eff}^{1/4}$, 
meaning 10d gravity is strong, dual to weak SYM, as usual). 

\subsection{Reheating model}

Now we remember that the cosmology we want corresponds to not just $N$ D2-branes, but the geometry will have a generic $CY_3$ 
replacing the transverse $S^6$, and D6-branes wrapping 4-dimensional surfaces $CY_2$ in this $CY_3$. 

Then the generic Euclidean string frame metric (Wick rotated using the domain wall/ cosmology correspondence) for this construction is of the type
\be
ds_{10}^2=d\vec{x}^2_3+dr^2+R^2(r)d\Omega(CY_3)^2\;,\label{10dmetric}
\ee
corresponding to a cone (which would be a "deformed flat space" if $R(r)=r$) with $CY_3$ replacing the sphere of the cone, somewhat analogous
to the Polchinski-Strassler construction \cite{Polchinski:2000uf} for $N$ branes,
with the sphere replaced by the $CY_3$, and Wick rotated on the radius $r$.

This background would have to be part of a solution of string theory. We also should have, {\em at $r=0$ only}, D6-branes wrapping 
$CY_{2,i}\subset CY_3$ surfaces, obtaining the $N_f$ chiral multiplets in the adjoint of $SU(N)$ as moduli of the $CY_3$ 
space and of the D6-branes (which are now 
fractional D2-branes, meaning they are wrapped on a space of vanishing radius, since the radius of the space is related to $r$, and 
$r\rightarrow 0$). Apart from this, we should also have a nontrivial dilaton and perhaps some nontrivial fluxes. 

Here, near $r=0$, $R(r)$ could be anything, modifying both the pre-geometric phase of the toy model, and the new geometric phase of the same. 
As an example of the above, consider the Maldacena-Nastase construction \cite{Maldacena:2001pb}
for fractional D2, obtained as NS5-branes wrapped on $S^3$, with flux and a twist. 
For it, in the 7 dimensional string frame solution $R(r)\sim r^{1/2}$ and $\phi=-r$ at infinity, and there is also a nontrivial flux. 
But in Einstein frame, due to the the fact that $e^{\phi/2}=e^{-r/2} $ multiplies $dr^2+R^2(r)d\Omega^2$, 
we find actually (since $dz^2=r^{r/2}dr^2$, so $z=4e^{r/4}$), 
$e^{r/2}R^2=e^{r/2}r$ = $z^2/4 \log z/4$, so $R(z)=z/2 \sqrt{\log z/4}$, meaning we have just a log correction to flat space. 
However note that this corresponds to a gravity 
dual to NS5-branes on $S^3$, giving a 3 dimensional theory for {\em apparent, or fractional} D2-branes only, so it is not good as a model. 

Once we have a solution in 10 dimensions near $r=0$, we compactify it on $CY_3$ 
to obtain a cosmology, as in the toy model case of the previous subsection. 
Under the compactification from $D$ to $d$ dimensions, $g_{\mu\nu}^{(d)}=\Delta^{\frac{2}{d-2}}g_{\mu\nu}^{(D)}$, and since now
$\Delta=R^6(r)$ is the volume and $d=4$, so $ds^2_4=ds_{10}^2\Delta$, after KK compactification we obtain
\be
ds^2_{E,4d}=R^6(r)d\vec{x}_3^2+R^6(r)dr^2.
\ee

Wick rotating and redefining the coordinates to obtain the cosmological ansatz
\be
ds^2_4=a^2(t)d\vec{x}^2_3-dt^2\;,
\ee
and assuming that we have a power law $R(r)\sim r^p$, we obtain
\bea
dt&=&R^3(r)dr\Rightarrow t\sim r^{3p+1}\Rightarrow r\sim t^{\frac{1}{3p+1}}\cr
a(t)&=&R^3\sim r^{3p}\sim t^{\frac{3p}{3p+1}}.\label{rat}
\eea

Comparing with the cosmological ansatz for a fluid with equation of state $p=w\rho$, we obtain
\be
a(t)\sim t^{\frac{2}{3(1+w)}}\Rightarrow 6p+2=9p(1+w)\Rightarrow w=-1+\frac{6p+2}{9p}\;,
\ee
or reversely, 
\be
p=\frac{2}{3(1+3w)}.
\ee

We note then that $R=r$, so $p=1$, implies the $w=-1/9$, not very good since it doesn't correspond to any reasonable matter. 
Also $R=r^{1/2}$, which means $p=1/2$, doesn't correspond to anything reasonable, since we obtain $w=1/9$. 
Reversely, RD cosmology, which has $w=1/3$, implies $p=1/3$, so $R(r)\sim r^{1/3}$, whereas Matter Dominated (MD) 
cosmology, which has $w=0$, 
implies $p=2/3$, so $R\sim r^{2/3}$. That means that 
the behaviour of the Maldacena-Nastase model, $R\sim r^{1/2}$, is right in between the RD and MD cosmologies. 

\subsubsection{S-NS5-branes at reheating time}

Finally, we come to the moment of reheating $t_0$, which after the Wick rotation corresponds to the position $r_0$. 
At the moment of reheating, we know from inflation (which is part of the general paradigm of holographic cosmology, albeit in a 
different corner, where there is no non-geometrical phase) that the energy in the scalar-gravity system is transferred to the Standard Model 
modes as radiation, thus "reheating" the cosmology. 

But that means that at least from this moment on, we need to have a string construction for the Standard Model, or more precisely, since 
we saw that we wanted to have ${\cal N}=1$ supersymmetry, of the MSSM. Since we had already considered wrapped D6-branes in this 
type IIA string theory construction for the holographic cosmology, it is reasonable to assume that we have an MSSM construction of intersecting 
D6-branes on the $CY_3$. In fact, this is one of the most popular type of constructions, and certainly the most relevant to the type IIA string 
theory case. 

Let us then further assume that the D6-branes are only created at time $t_0$, or position $r_0$, and see if this can be made consistent. 
How can we create a D6-brane? In the $r$ picture, the D6-brane must end on a brane, and the only possibility is an NS5-brane. 
\footnote{Indeed, since the fundamental string, or F1-brane, can end on a D0-brane (or D-particle), meaning there are 
Dirichlet boundary conditions in all directions, {\em except time}, 
it means that in the direction $x$ parallel to the string (we can choose coordinates in which F1 is parallel with $x$), 
the D0 (D-particle) ends it at $x=x_0$. 
This means that the field strength $F_{\mu\nu}=F_{(2)}=dA_{(1)}$ on the D0-brane ($F_{(2)}$ is 
electric/magnetic dual to $F_{(8)}=dA_{(7)}$) gives a charge for $B_{\mu\nu}=B_{(2)}$ on the 
string.  After a Poincar\'{e} duality (a generalization of Maxwell duality 
for antisymmetric $p+1$-form fields) on this configuration,  $H_{(7)}=dB_{(6)}$ 
(which gives a magnetic charge for $B_{\mu\nu}=B_{(2)}$) is an electric charge for $A_{(7)}$ on the D6-brane, so we see that 
indeed a D6-brane can end on an NS5-brane.}
Here there are 5+1 directions common to both the NS5-brane and the D6-brane, and one direction $x$ that is now parallel to the D6-brane, 
but ends at $x=x_0$, where is located the the NS5-brane. 
This is a spatial direction, and time is common to both, but by the same double Wick rotation needed for holographic cosmology, 
we find that at $r_0=t_0$, which is the holographic time of reheating, we create a D6-brane, but with the same spatial worldvolume, 
both in our 3 dimensional one and in the $\Sigma_3\subset CY_3$. The NS5-brane is a spatial, or S-brane, as defined in 
\cite{Gutperle:2002ai} (in that paper, S-D-branes were defined in detail, and S-NS5-branes less so, but they are consistent).

Also note that this is the only way we can create intersecting branes at a given holographic time $r_0$, since D4-branes 
don't end in some 3-brane, and D8-brane also don't end in some 7-brane (since we don't have these objects in type IIA string theory), 
meaning the MSSM {\em must} be obtained from intersecting D6-branes, created from corresponding NS5-branes at
$r_0$. That means that within this type IIA holographic cosmology construction, our construction for reheating is unique. 

Since each of the D6-branes ends on an NS5-brane, and we have intersecting D6-branes in complementary parts of the $CY_3$
space, we have also NS5-branes intersecting in the same way, and de facto covering all of the $CY_3$. 
\footnote{We can also add more NS5-branes than needed to create the intersecting D6-branes, 
to make a distribution of NS5-branes on the $CY_3$ that covers the whole of $CY_3$. The D2-branes and fractional D2-branes 
generate a flux through the transverse $CY_3$, that should stabilize the gravitational force that would want to collapse 
the "$CY_3$-spherical" distribution of mass at $r=r_0$ (like a soap bubble stabilized from collapsing by the air pressure inside).
In that case, the NS5-brane flux of the NS5-brane distribution would saturate the {\em remaining} 
flux of the D2-branes and fractional D2-branes, not saturated at $r=0$, while 
the flux of the few extra ones is saturated by creating the intersecting D6-branes. }

What about the form of the cosmological solution, with a certain $R(r)$ leading to a certain $a(t)$? We saw that the Maldacena-Nastase 
solution, for instance, does not obtain the RD cosmology. But we shouldn't expect it to, in fact. 
Around the NS5-branes at $r_0$, the form of $R(r)$ can be governed by 
the gravitational solution of D2-branes and NS5-branes, but after that (at $r>r_0$), 
{\em if the charge of the D2-branes is completely screened by the NS5-branes}, 
we have flat-like space (flat space plus extra dimensional $CY_3$), meaning that the  cosmological 
solution is governed by the usual calculations in hot Big Bang, with radiation corresponding to MSSM excitations on the
intersecting D6-branes, and we get RD cosmology in the usual way, from the FLRW  equations. 
Note that we can have a small flux that is still unscreened by the NS5-branes, that will not interfere with the flatness of the space, 
since only a large number $N$ of flux will interfere, as in the $AdS_5\times S^5$ example, for instance. 

But how do we get these MSSM excitations giving radiation, and what is their temperature (the reheat temperature)? 
The calculation for the second question (reheat temperature) is related to the first, and for comparison we will also consider the 
standard, perturbative, one in the section 5 (given that in standard reheating computer simulations give a result not much 
different than the perturbative one). 

To understand the first question, we consider the recent calculation of reheating due to an S-brane in \cite{Brandenberger:2020wha} 
(following \cite{Brandenberger:2020tcr,Brandenberger:2020eyf}). There, reheating is due to an S-D2-brane (3 spatial coordinates, 
namely our own) located at some $t=t_0$, and modelled as a tachyon condensate  $T(x,t)$ ("vortex", but in $x$ and $t$ instead of 
$x$ and $y$) on an unstable D4-brane (worldvolume $t,x,y,z$, together with an extra coordinate $\phi$). This is the usual construction 
for unstable D$p+2\rightarrow$ D$p$, just with the D$p$ being an S-brane. The tachyon potential is 
\be
V(T)=-\frac{\lambda}{2}\eta^2T^2+\frac{\lambda}{2}\eta^4\;,\;\;\;|T|\leq \eta\;,
\ee
and $V=0$ for $|T|>\eta$. In their case, the cosmology was of ekpyrotic type, with a potential for the scalar corresponding to the extra coordinate
$\phi$ of the type $V(\phi)=-V_0e^{\sqrt{\frac{2}{p}}\frac{\phi}{M_{\rm pl}}}$, and they find reheating occurs with a final radiation density of 
\be
\rho_A\sim \frac{3}{16\pi}\lambda V_0\;,
\ee
so reheating is very efficient if we have the generic $\lambda\sim 1$ (remember that $\lambda$ is a coupling parameter for the potential to 
create an S-brane, and $V_0$ is the scale of the potential for the cosmology). 

The reheating mechanism is as follows. The radiation that results corresponds to Standard Model 
gauge fields $A_\mu$ like electromagnetism, on the worldvolume of 
an effective D3-brane (or D$p$-brane wrapped on constant $p-3$ cycles in the extra dimensions), with action 
\be
S=\int dtd^3x V_{\rm total}(T) \sqrt{\det(\eta_{\mu\nu}+F_{\mu\nu})}\;,
\ee
giving the coupling of the S-brane to radiation. The S-brane is assumed to form when the total energy density, $V(\phi)$ plus kinetic, becomes close 
to the energy density (tension) of the S-brane. 

To summarize, we see that the energy is initially in the usual gravity plus scalar $\phi$ modes (like in inflation), 
then is transferred into the S-brane, which then gives 
radiation ((MS)SM modes), the transfer being governed by the tachyon potential $V(T)$, and the process is non-perturbative.

We note that the mechanism can be exactly imported into our case, even though the cosmology is different. The different cosmology is just related
to a different potential $V(\phi)$, but the mechanism is the same. In our case we also have Standard Model gauge fields on the intersecting 
D6-branes, giving effective D3-branes, that can interact with the S-brane. 
While they have S-D2-branes, viewed as vortices on D4-branes, we have S-NS5-branes, which are harder
to describe explicitly, but their construction is related to a similar case via a chain of dualities. Indeed, we can imagine 2 extra compact dimensions
wrapped by the branes in \cite{Brandenberger:2020wha}, so that actually we have S-D4-branes as vortices on D6-branes, and the 
mechanism and calculation would remain the same. Then, a T-duality on a commonly transverse extra dimension relates them to S-D5-branes 
as vortices on D7-branes, then an S-duality relates them to S-NS5-branes as vortices on D7-branes, and finally another T-duality on a commonly 
transverse extra dimension to S-NS5-branes as "vortices" on D8-branes. 

It is not completely clear if the calculation of the energy can  
be imported, both because of the chain of dualities, and because of the different potential $V(\phi)$, but
the fact that the reheating is efficient can be. We can then infer that the density of the resulting radiation will be a fraction not much smaller than 1 
of the total energy before the reheating time $t_0$. If we import also the actual coefficient and define a corresponding reheat temperature, we 
get
\be
T_{RH}^4\equiv \rho_A\sim \frac{3}{16\pi}\lambda V_0.\label{TRH4}
\ee

Here $V_0$, which is taken to mean the potential or energy density at the time $t_0$ of creating the S-NS5 (or $r_0$ in the 
radial time picture) 
must be related to the scale at which we create the S-NS5-brane in cosmology. In field theory, the NS5-brane is created some 
time after reaching the string energy scale, $q=U_{\a'}=1/\sqrt{\a'}$. In cosmology, this is some time after the 1 in $H_2(r)$ becomes dominant, 
so when the energy scale becomes somewhat smaller than the string scale, when $V=V_0<\a'^{-2}$. 
Then with $\lambda\sim 1$, we find $T_{RH}^4<(3/16\pi)\a'^{-2}$. 

In section 5, we will consider, as an alternative, the perturbative calculation for the reheating temperature, since we know that usually, 
computer simulations find a result close to the perturbative one. Here we will also find that the result is close to the above (parametrically). 

\section{The view from the 2+1 dimensional field theory}

Since we are constructing a holographic cosmology, we should be able to describe everything from the point of view 
of a 3-dimensional Euclidean field theory. In this section, we do that.

We have seen how to describe the pure gravity + scalar mode (the analog of inflation) in a non-geometric phase as a 3 dimensional Euclidean
field theory for adjoint fields. We need to understand then how to add the Standard Model (or rather, MSSM) modes from the point of view of the 
3 dimensional Euclidean theory.

\subsection{The general method}

The Standard Model modes in cosmology are modes on the intersection of D6-branes. 
The general method for adding something that corresponds to branes wrapping in cycles in the gravity dual can be explained from 
standard case of the ${\cal N}=2$  superconformal field theory (SCFT) with $SO(8)$ symmetry
obtained from the orientifolding of D3-branes and D7-branes (branes at the orientifold 
fixed point), with gravity dual the orientifold of $AdS_5\times S^5$ \cite{Aharony:1998xz}. 

In this case we have $N$ D3-branes and 4 D7-branes, identified under the orientifold, with fixed O7-plane, leading to an $SO(8)$ global symmetry. 
The global symmetry arises in the decoupling limit for the D7-branes, decoupling the $SO(8)$ gauge fields corresponding to strings extending
between two D7-branes.
The strings extending between 2 D3-branes at the fixed plane give rise to an $USp(2N)$ gauge fields, the strings between the D3-branes 
and D7-branes lead to bi-fundamental matter, which in the decoupling limit is just fundamental matter for the $USp(2N)$ gauge group, 
with global $SO(8)$ symmetry. The model has ${\cal N}=2$ supersymmetry because the orientifolding breaks half of the supersymmetry. 

In the gravity dual, we start with the $AdS_5\times S^5$ gravity dual for the $N$ D3-branes, and perform the orientifolding with fixed O7-plane, 
which corresponds to an worldvolume of $AdS_5$ times an $S^3\subset S^5$. 
In other words, one performs the orientifolding only on $S^5$, leaving the $S^3$ invariant. One then {\em adds } 4 D7-branes on top of the O7, 
in the decoupling limit, i.e., adds an $SO(8)$ SYM vector multiplet in the 7+1 dimensional space corresponding to the O7-plane, $AdS_5\times 
S^3$. So the "gravity dual" is not just gravity, but is 10 dimensional gravity plus 8 dimensional $SO(8)$ vector multiplet, which can be 
KK expanded on the $S^3$. The resulting KK modes in 4+1 dimensions (in $AdS_5$), with $SO(8)$ gauge symmetry (and $SO(4)$ 
symmetry from the KK expansion) act as sources for $USp(2N)$-gauge invariant operators of the 3+1 dimensional SCFT, with 
$SO(8)$ (and $SO(4)$) global charges. 

We see the method emerging: gauge and matter fields for $SO(8)$ in $AdS_5$ (coupled to gravity) are gravity/gauge dual to fundamental matter with
global $SO(8)$ in the field theory. We can build a more complete general picture based on the above example, see for instance 
\cite{Nastase:2015wjb}.

We also note that the SYM KK modes covered all of $AdS_5$, but we could imagine creating them at a point $r_0$ in the radial direction
(which in the case of holographic cosmology is Wick rotated time $t_0$). That would correspond to creating, or rather adding in, the fundamental 
matter in the dual field theory at some momentum scale $q_0$ along the inverse RG flow. That is certainly consistent, since along the RG flow 
one averages over (integrates out) modes, so along the inverse RG flow one can add in modes.

\subsection{The gauge theory}

Armed with this understanding, we consider our concrete case, of creating matter with $SU(3)\times SU(2)\times U(1)$ gauge group, 
coming from intersecting D6-branes on 3-surfaces in $CY_3$, at time $t_0$, corresponding to $r=r_0$ radial direction. 

In the field theory, this corresponds to adding in, at some (inverse RG flow) momentum scale $q_0$, 
extra $SU(N)$ fundamental matter fields, besides the adjoint $A_i^a$, $N_s$ 
scalars $\Phi^{aM}$ and $N_f$ fermions $\psi^{aL}$ that are dual to gravity plus scalar. As in the D3-D7 brane case above, 
the gauge group $SU(3)\times SU(2)\times U(1)$ in the gravity dual
becomes a global symmetry group for the 3 dimensional fundamental matter to be added. 

Defining generically $m$ as being a fundamental index in $SU(3)\times SU(2)\times U(1)$, with $\bar mn$ as the adjoint group in the same, 
the Standard Model (or rather, MSSM) fields will be denoted generically $A_\mu^{\bar m n}$ for gauge fields, $Q^m$ for quarks (and fermionic 
superpartners) and  $H^m$ for Higgs (and squarks). 

Then the matter added in is: scalars $q_u^m$ and fermions $\chi_u^m$, where $m$ is the generic fundamental index
of $SU(3)\times SU(2)\times U(1)$, and $u=1,..,N$ is the fundamental index of $SU(N)$ (so that the adjoint $a=\bar u v$, bifundamental), 
{\em a scalar in 3 dimensions for each gravity dual (MSSM) scalar and a fermion in 3 dimensions for each gravity dual (MSSM) fermion.}

The gauge invariant operators ("currents") dual to the Standard Model (or MSSM) fields in the gravity dual are ${\cal O}_1^m$ dual to 
quarks $Q^m$, the vector operator (global symmetry
current) $J_\mu^{\bar m n}$ dual to the gauge fields $A_\mu^{\bar m n}$ ($\bar m n$ is an adjoint 
index of MSSM) and the scalar operator ${\cal O}_2^m$ dual to the Higgs doublet field $H^m$. 

Then, formally
\bea
J_\mu^{\bar mn}&\sim& \bar \chi_u^{\bar m} \gamma_\mu\chi_u^n + \bar q_u^{\bar m} \d_\mu q_u^n\cr
{\cal O}_1^{m'}&\sim& q_u^{\bar m} \chi_u^n M_{\bar mn}^{m'}\cr
{\cal O}_2^{m'}&\sim& (q_u^{\bar m} q_u^n+\chi_u^{\bar m} \chi_u^n)M_{\bar mn}^{m'}\;,
\eea
where the matrix $M_{\bar mn}^{m'}$ are Clebsch-Gordan coefficients turning 2 representations into 
another.

As we see, the addition of these fields dual to Standard Model fields would not normally change the gravity dual, since they are much fewer than 
the scalars and fermions that construct the gravity dual (which are $N_s\simeq N_f\sim 10^4$). Except that when we add them in, 
we are at a momentum scale 
past the scale $q=1/\sqrt{\a'}$ at which string corrections come in, and outside the decoupling limit anyway. In that case, the 
field theory is not anymore weakly coupled due to small $g^2_{\rm eff}=g^2_{YM}N/q$, but rather strongly coupled due to large 
$\tilde\lambda_{\rm eff}=\a' q^2$, so the field theory picture for reheating and beyond (i.e, RD cosmology) 
is not useful or even possible, and we must be content with the gravity picture from the previous section.

\subsection{Comments on the brane construction for the field theory}

In the standard case of D3-D7-O7, we saw that the field theory corresponded to some brane construction ($N$ D3-branes and 4 D7-branes 
at the O7 planes). We also saw that the brane construction for the field theory was simply related to the brane construction in the gravity 
dual: $N$ D3-branes plus O7 plane make the gravity part of the gravity dual, and the D7-branes in the field theory construction 
(at the O7 plane) correspond to D7-branes in the gravity dual (at the O7 plane). 

In our case, the intersecting D6-brane construction in the gravity dual (cosmology) was not defined, and also the $D6_i$ branes giving the adjoint 
scalars and fermions were not specified. That means that we cannot define the brane construction for the field theory. 

But we can make the same observation: the D6-branes in the gravity dual (cosmology) correspond to D6-branes making up the field theory, 
both for the intersecting ones (MSSM) and for the $D6_i$ ones.

\section{Particle phenomenology and the cosmological constant}

We saw that reheating of holographic cosmology was defined in the context of an MSSM construction, so we
might ask if we cannot learn something about particle phenomenology. Moreover, as explained in \cite{Nastase:2018cbf} (with the 
caveat that there was no reheating model there), 
the cosmological constant is expected to evolve with the field theory momentum $q$ as a decreasing power law, 
so we might ask whether now we can understand more precisely the current low cosmological constant.

\subsection{The cosmological constant and supersymmetry breaking}

\subsubsection{The cosmological constant from field theory}

The first thing to do is to define the cosmological constant from the point of view of the 3 dimensional field theory. 

The scalar inflaton $\phi$ in cosmology (in the gravity dual) corresponds in field theory to a scalar operator. Together with 
the spatial (3 dimensional) part of the cosmological metric, $g_{ij}$, the inflaton $\phi$ is dual to the energy momentum tensor 
of the field theory, $T_{ij}$. As we mentioned,  cosmological evolution in radial direction $r$, Wick rotated to cosmological time $t$, 
is dual to the inverse RG flow. 

Then the scalar potential in cosmology, including the value of the cosmological constant $\Lambda$, implying
the value of the radius of (A)dS  cosmology, is dual to the central charge $c$ of the 3 dimensional field theory, or 
more generally the value of the c-function. We now calculate the relation between the two. 

For $AdS_{d+1}$, the cosmological constant is related to the AdS radius by
\be
\Lambda=-\frac{d(d-1)}{2R^2}[M_{\rm Pl, d+1}]^{d-1}\;,
\ee
so in particular for $AdS_4$, we have 
\be
\Lambda=-3\frac{M_{\rm Pl, 4}^2}{R^2}=-\frac{3}{8\pi G_{N,4}R^2}.
\ee

Also for $AdS_{d+1}$, the dual central charge is 
\be
c=\frac{\pi^{d/2}R^{d-1}}{\Gamma(d/2)[l_{\rm Pl,d+1}]^{d-1}}=\frac{\pi^{\frac{d-2}{2}}R^{d-1}}{\Gamma(d/2)8G_{N,d+1}}\;,
\ee
so in particular for $AdS_4$, we have
\be
c=\frac{R^2}{4G_{N,4}}.
\ee

Then, for $AdS_4$, we have the relation between cosmological constant and central charge,
\be
\Lambda =-\frac{3c}{2\pi R^4}=-\frac{3}{32\pi G_{N,4}^2c}=-\frac{6\pi M_{\rm Pl,4}^2}{c}\;,\label{lambdasign}
\ee
meaning when $c\rightarrow \infty$, $\Lambda\rightarrow 0$. But we want to find a more general relation, in the case of a scalar potential, 
which corresponds to an RG flow, with a c-function.

In this case of a c-function in 3 dimensional field theory, dual to the RG flow-type gravity dual metric (domain wall)
\be
ds_{d+1}^2=e^{2A(y)}d\vec{x}_d^2+dy^2\;,\label{RGflow}
\ee
we find the generalized relation
\be
c(y)=\frac{\pi^{d/2}}{\Gamma(d/2)[l_{\rm Pl,d+1}A']^{d-1}}.
\ee

For $d=3$, so for $AdS_4$, the RG flow-type metric (domain wall) is
\be
ds_4^2=e^{2A(y)}d\vec{x}_3^2+dy^2
\ee
and is Wick rotated by the domain wall/cosmology correspondence to
\be
ds_4^2=e^{2A(t)}d\vec{x}_3^2-dt^2=a^2(t)d\vec{x}_3^2-dt^2.
\ee

The Hubble scale of this cosmology is
\be
H=\frac{\dot a}{a}\rightarrow A'\;,
\ee
where we also wrote the corresponding value in the domain wall,
so the central charge if its dual field theory becomes
\be
c(t)=\frac{1}{4G_{N,4}[A'(r)]^2}=\frac{1}{4G_{N,4}H^2}=\frac{2\pi M^2_{\rm Pl,4}}{H^2}.\label{cH}
\ee

Finally, the cosmological constant is related to the Hubble scale by
\be
\Lambda =3H^2M^2_{\rm Pl,4}\;,
\ee
and by the Friedmann equation corresponds to the potential $V(\phi)$, in the RG flow situation. 
Then we find 
\be
c(t)=\frac{6\pi M^4_{\rm Pl,4}}{V(\phi(t))}.\label{cV}
\ee

Note that in going from (\ref{lambdasign}) to the above, we dropped a sign, in effect continuing between the AdS and dS cases. 
But at the least in the domain wall case, that makes sense, since the template for that is the case of $N$ D2-branes, for which $c$ is well 
defined, and then we made the double Wick rotation called domain wall/cosmology correspondence to this case with $V>0$. 
The only issue is the limit when $V$ is constant and equal to an exactly constant $\Lambda$, but we assume nothing drastic happens 
in the limit. 

Then the rapid evolution of the scalar in the scalar potential, decreasing the potential $V$ and creating energy to be transmitted to the (MS)SM, 
is mapped in field theory to the rapid increase of the central charge $c$ {\em of the adjoint modes} (dual to gravity)
along the inverse RG flow. Note that also the {\em total} number of degrees of freedom, i.e., $c_{\rm total}$, and the 
(MS)SM number of degrees of freedom, so also $c_{SM}$ (related to the matter entropy), {\em increases} along the inverse RG flow. 

That means in the field theory dual to reheating that the rapid increase
in $c$ and the coupling of gravity to matter imply that some central charge $c$ can be
transferred from adjoint modes to the MSSM degrees of freedom. 

Note that we will argue in subsection 5.1.3 that relations (\ref{cH}) and (\ref{cV}), the relation to between the central charge and 
potential, or cosmological constant, is true only just until the end of reheating, when the number of modes and $c$ is fixed, resulting in a 
fixed $V(\phi)=\Lambda$, a {\em true} cosmological constant, that remains constant until today. 

\subsubsection{The reheat temperature}

We now find another way to estimate the reheating temperature, which for inflation is related to 
the Hubble scale of inflation, $H_{\rm infl}$, and the coupling of gravity modes to MSSM modes, namely the perturbative 
calculation (which usually is a good enough estimate, even for the nonperturbative case). Here as usual, the gravity + scalar 
mode is coupled to the MSSM modes, according to the mechanism in  \cite{Brandenberger:2020wha}, essentially a nonperturbative 
coupling of gravity + scalar to S-brane, then S-brane to radiation, as described in section 3, but we use standard perturbative 
theory instead of the calculation done there, and find that the result is nevertheless not much different.

For reheating where the boson decays in fermions, $\phi\rightarrow \psi\psi$, we have, from standard inflationary perturbative reheating theory, 
also applicable here,
\be
T_{RH}\sim \frac{1}{g_*^{1/4}}\frac{g}{\sqrt{8\pi}}\sqrt{m_\phi M_{\rm Pl}}\;,
\ee
where $g$ is the Yukawa coupling, $m_\phi$ the boson mass and $g_*$ the number of modes.

If we reheat at a field theory energy scale of the order of the string scale, $U=q\sim 1/\sqrt{\a'}$, (corresponding to a distance 
in the gravity dual $r\sim \sqrt{\a'}$), it means we reach {\em string } modes, so there is a {\em drastic} increase in 
the effective field theory central charge $c$, so the potential $V(\phi)$ of the cosmology decreases drastically, as expected in (p)reheating. 

We can relate the Yukawa coupling and boson mass in cosmology to the string coupling $g_s$ as
\bea
g&\sim &  \frac{\kappa_{N,4}}{\sqrt{\a'}}\propto g_s \sim e^\phi\cr
m_\phi&\sim & \frac{1}{\sqrt{\a'}}\propto \frac{g_s}{\kappa_{N,4}}\;,
\eea
implying for the reheat temperature
\be
T_{RH}\sim \frac{1}{g_*^{1/4}}\frac{g_s^{3/4}}{\sqrt{8\pi}}M_{\rm Pl}.
\ee


After reheating, we must transition to RD cosmology, with $c$ eventually dominated by MSSM modes, not gravity 
modes.

Assuming that, even after reheating, we still have $\Lambda \propto 1/c$, and moreover that in the RD Universe 
we have gravitational cosmological holography, meaning that
the number of degrees of freedom is (less than, but almost equal to) 
the area of the cosmological horizon volume divided by $L_{\rm Pl}^2$, so 
\be
\frac{M_{\rm Pl}^4}{\Lambda}\sim c\sim N\lsim \frac{H^{-2}}{L_{\rm Pl}^2}=\frac{M_{\rm Pl}^2 }{H^2}\Rightarrow \Lambda\lsim H^2 M_{\rm Pl}^2\;,
\ee
which is consistent with the standard inflationary formula (from the Friedmann equation) $\Lambda =3H^2M^2_{\rm Pl,4}$.
Since we obtain a correct result, we can infer that the initial assumptions were also correct.

Based on the experimental evidence (suggesting the cosmological constant is a true constant, not a quintessence), we
expect that {\em the cosmological constant remains constant after reheating}, once all the fundamental degrees of freedom have been 
accessed, both in field theory and in string theory. This is justified theoretically by the fact that the gravity solution (the cosmology) 
has the harmonic function $H_p$ reducing to one before reheating, so it has no
dependence on $g^2_{YM}N$, so no flowing with  the field theory energy scale $U$, dual to cosmological time $t$, anymore. 
Moreover, for times $t>t_0$, meaning in field theory energies $U>U_0$, there are no more degrees of freedom "added in"
(when going to the IR, we lose degrees of freedom, when going to the UV, we add them), so $c$ is constant, and then by the 
formula from the previous subsection, $\Lambda$ is also constant. Note also that experimentally, a (true) cosmological constant 
is better than quintessence (varying potential), in particular it solves better the Hubble tension (tension in measurement of $H_0$ 
at different scales), see \cite{Banerjee:2020xcn}.

In the toy model of $N$ D2-branes, and assuming that the formula (\ref{ND2}), valid in the non-geometric phase only (in the harmonic 
function $H_p$ neglect the 1, opposite to the reheat case where $H_p\simeq 1$), we obtain
\be
T_{RH}\sim \frac{e^{3\phi/4}}{g_*^{1/4} \sqrt{8\pi}}M_{\rm Pl}\sim \frac{1}{N}\frac{\lambda_{\rm eff, RH}^{15/16}}{g_*^{1/4}\sqrt{8\pi}}M_{\rm Pl}.
\ee
Here we have assumed that reheating is close to the point where in $H_p$ the 1 and the term with $\lambda_{\rm eff}=g^2_{YM}N/U$ 
are equal, so we can still use the formula (\ref{ND2}) at reheating, up to a factor of order 1. But then $\lambda_{\rm eff, RH}\sim 1$ 
(parametrically, i.e., not much smaller than 1), so we find $T_{RH}^4\sim M_{\rm Pl}^4/[N^4g_*(8\pi)^2]$. On the other hand, from the same
(\ref{ND2}), $g_{s,RH}\sim 1/N$, and $1/\sqrt{\a'}\sim g_s M_{\rm Pl}$ (all formulas true only parametrically), so we obtain 
$T_{RH}^4\sim  M^4_{\rm Pl}/[g_*N^4(8\pi)^2]$, not too different from the (imported) S-brane result from section 3 (\ref{TRH4}), 
which was $T_{RH}^4\lsim (3/16\pi)\a'^{-2}\sim 3g_{s,RH}^4M_{\rm Pl}^4/[16\pi]\sim 3M^4_{\rm Pl}/[(16\pi)N^4]$.

Then we see that the reheating temperature
can be made small with respect to $M_{\rm Pl}$, though it will always be much larger than the BBN temperature, as is 
generically required. Phenomenologically (from the fit to CMBR data), we saw that $N\sim 3000$, 
and $\lambda_{\rm eff, RH}$ can be made small, though not too small: in fact, when we remember that in $H_p$ we used the $\lambda_{\rm eff}$
term, when we should have used the 1 term, it is clear the value is not too small, it is still consistent with 1, as we put above for comparison 
with the formula from the section 3 (\ref{TRH4}). 
Also, the above formula was for the toy model, but in reality, the extra $D6_i$ branes will modify it.



\subsubsection{Matching the cosmological constant against experimental data}

We should see if it is possible to obtain a cosmological constant that is as small as the one measured experimentally. 

{\em First iteration}.

Just before reheating, the cosmology is described by perturbative 3 dimensional SYM, 
dual to strongly coupled gravity, or more precisely string theory, not too different from the 
inflationary type, meaning, $V=\Lambda$ almost constant or slowly varying, so $H\simeq$ almost constant constant and $c$ 
almost constant.  True string theory, coupling the two sectors, SYM and gravity, starts then (shortly before reheating). 
String theory modes are accessed, and $c$ grows tremendously, so $V$ drops to the minuscule $\Lambda$ that we have now, which is 
truly constant, not just a quintessence, as we argued above. 
At reheating and afterwards, we have not only coupling of SYM to gravity, 
but also of gravity to MSSM modes, related to the dropping energy of the gravity side, i.e., dropping potential $V$.

Then, the adjoint field theory central charge $c_{SYM}$, dual to the 4 dimensional gravity modes, and now also coupled to them, and now 
also coupled to the 4 dimensional MSSM modes,  or their 3 dimensional dual, the fundamental field theory modes, 
start being transferred to MSSM modes. 
Reheating ends when most of the available cosmological energy is transferred into MSSM modes, 
and equipartition of modes happens, so we expect $c_{MSSM}\simeq c_{\rm gravity}$. 

{\em Assuming at this point we still have the inverse relation between central charge and $\Lambda$}, and considering central charge in 
cosmology as a 
stand in for entropy ($S\sim \ln N_{\rm states}\sim \ln N_1^{n_{dof}}\sim n_{dof}\ln N_1$, where $n_{dof}$ is number of degrees of 
freedom, $N_1$ is the number of states per degree of freedom), as well as the fact that around reheating, the field theory and gravity 
(and also MSSM) are coupled, so equipartition is expected, we have 
\be
(S\sim c_{MSSM}\sim c_{\rm gravity})\propto c_{\rm field\; theory}\propto \frac{M_{\rm Pl}^4}{\Lambda}\;,\label{SLambda}
\ee
which defines $\Lambda$ from the cosmological entropy $S$, but from experimental data on $S$,
we obtain a $\Lambda$  that is much larger.

Indeed, we know that 
at the end of reheating we want to end up with an entropy of $S\gsim 10^{88}$ (see for instance eq. (10.35) in \cite{Nastase:2019mhe}), 
which means that for consistency with the above formula, we must start off, before reheating, with a $\Lambda \sim 10^{-35} M_{\rm Pl}^4$, 
or at an (potential) energy scale (in the gravity dual theory) of $M\sim 10^{-9} M_{\rm Pl}\sim 10^{10} GeV$, yet with a negligible
cosmological entropy $S\sim 1$ (or not too much bigger than 1, anyway). Note that the large entropy at the end of reheating is understood 
in cosmology as being the result of the efficient transfer of energy (see subsection 3.3.1) into the MSSM modes, due to the coupling of the 
S-NS5-brane to the D6-branes, resulting in a large $T_{\rm RH}$ and large $S$. 

But in reality, when reaching the string scale in field theory, at $U=1/\sqrt{\a'}$, at which time in cosmology we have
some cosmological constant, due to the mechanism described in \cite{Nastase:2018cbf} (inverse RG flow leads to dropping 
of $\Lambda$), the cosmological entropy is negligible still (in some sense $S\sim 1$). We assume that the same $c_{\rm field \; theory}
\propto 1/\Lambda$ relation still holds for $U>1/\sqrt{\a'}$, yet we require consistency (equality of the two relations) at $U=1/\sqrt{\a'}$.

This means that we must have that the observed cosmological entropy, after the string scale, 
is actually smaller by the ratio of the cosmological 
constant and Planck scale at the end of the non-geometrical phase, and equal to
\be
(S_{\rm obs.}\sim c_{MSSM, obs.}\sim c_{\rm gravity,obs.})\propto\frac{\Lambda_{\a'}}{M_{\rm Pl,4}^4}c_{\rm field\;theory}
\propto \frac{\Lambda_{\a'}}{\Lambda}\;,
\ee
where $\Lambda_{\a'}$ is the cosmological constant corresponding to reaching the string scale in the dual 3 dimensional field theory, 
and we see that it replaces $M_{\rm Pl}^4$ in (\ref{SLambda}). Then 
this cosmological constant  $\Lambda_{\a'}$ corresponds to what was
analyzed in \cite{Nastase:2018cbf}. There we did not have a model of reheating, so the true $\Lambda$ arises now. 

Moreover, we see that the entropy is mostly created during the reheating, but it really happens because as we go to the 
UV in field theory, we have more degrees of freedom (the arrow of time corresponds to this universal property of RG flows), 
as argued in \cite{Nastase:2019rsn,HNSoon}. 

{\em Second iteration}

We see that we need to find a way to obtain a cosmological constant of $10^{-35}M_{\rm Pl}^4$, or a potential energy scale of 
$M\sim 10^{-9}M_{\rm Pl}$ just before reheating, in order to obtain consistency with experimental data. 

Considering the KK compactification of string theory gravity from 10 to 4 dimensions (particle physics, MSSM modes, are at the intersection of 
D6-branes, but gravity is still 10 dimensional, and with a compact $CY_3$), we have 
\be
M^2_{\rm Pl,4d}=M^2_{\rm Pl,10d} V(CY_3)\;,
\ee
where $V(CY_3)$ is the dimensionless volume of the manifold (taking out the dimensionful scale). Moreover, we obtain the relation between 
the string and 4 dimensional Planck scales
\be
\frac{M_{\rm Pl,10d}^8}{2}=\frac{1}{(2\pi)^7g_s^2\a'^4}\Rightarrow  \frac{1}{\sqrt{\a'}}
\sim 10 e^{\phi/4}M_{\rm Pl,10d}=10 e^{\phi/4} \frac{M_{\rm Pl,4d}}{\sqrt{V(CY_3)}}\;,\label{strings}
\ee
where we have approximated the numerical factors.

In the case of RD cosmology, starting at reheating, we have $a(t)\sim t^{1/2}\sim r$ from (\ref{rat}) and $r\gg \sqrt{\a'}$.
Then at reheating, we start at an energy 
\be
E\sim a(t)^{-1}\sim r^{-1}\ll \frac{1}{\sqrt{\a'}}\sim g_s^{1/4}M_{\rm Pl,4d}/\sqrt{V(CY_3)}\;,\label{lambdaE}
\ee
which means, using (\ref{UUin}), that
\be
\Lambda_{\a'} \sim E^4\ll \frac{g_s}{[V(CY_3)]^2}M^4_{\rm Pl,4d}\sim \frac{1}{N}\frac{U_{\rm initial}}{U_{\a'}}
\frac{1}{[V(CY_3)]^2}M^4_{\rm Pl,4d}.\label{lambdaa}
\ee

We must also have $g_s$ extremely small, since $g_sN$ must be $\ll 1$ (by the amount
of RG flowing in (\ref{rat})) and $N$ large, and we can easily choose $V(CY_3)$ to be large. 
Note that because of (\ref{range}), we have $U_{\a'}/U_{\rm initial}\gg N^{4/5}$. Let's say $U_{\a'}/U_{\rm initial}=N$, so 
$g_s\simeq 1/N^2\sim 10^{-7}$ from the phenomenological fit. However, that is not much help in going from $\Lambda/M_{\rm Pl}^4$ of 
$10^{-88}$ to $10^{-120}$, we only reach $10^{-95}$ by using $g_s\sim 10^{-7}$ in (\ref{lambdaa}). 

We still have 25 orders of magnitude left, but then choosing $V[CY_3]\simeq 10^{12}\simeq (100)^6$, we can fix the remaining difference 
in $\Lambda_{\a'}$ in (\ref{lambdaa}). This very reasonably small dimensionless volume of the $CY_3$ could be understood as being due to 
a radial direction (corresponding to Wick rotated time)
$r\simeq 100 \sqrt{\a'}$ when we start with the RD cosmology relation (see (\ref{rat}) 
$R(r)\sim r^{1/3}$, i.e. $R(r)\simeq \sqrt{\a'} (r/100\sqrt{\a'})^{1/3}$ or so.\footnote{Note 
that this seems to be suggest $R\simeq r$ at this scale on the average over the $CY_3$, but
it might not necessarily be like this; it could be that it is just a coincidence, or that there is something at that scale that 
changes the $R\simeq r$ to $R\simeq r^{1/3}$, etc.} 

Finally then, after reheating, we have a consmological constant
\be
\Lambda\equiv E_0^4\label{LambdaE0}
\ee
that is truly constant (doesn't evolve anymore), but its energy scale, $E_0$, becomes a geometrical quantity again (like $E$ related to
$\Lambda_{\a'}$ in (\ref{lambdaa}) was before reheating (\ref{lambdaE})) only now, when $\Lambda$ dominates the cosmology. 

The above was a reasonable scenario for entropy generation. But then we must also presumably have some MSSM entropy 
produced during our non-geometric ("inflationary") period, though perhaps now we don't need to, 
since we have a small coupling to MSSM during this period, and we 
just create entropy through the strongly coupled gravity at the starting region of the reheating. 

Note that we have used the boundary condition for the cosmological constant that $\Lambda=M_{\rm Pl}^4$ if $g^2_{\rm eff}=1$, since this is the 
overlap domain for the gravity dual and the SYM, and then we can assume that $c_{SYM}\sim 1$ at this point, and then 
the relation with $c\sim M^4_{\rm Pl}/\Lambda$ implies $\Lambda \sim M^4_{\rm Pl}$.

\subsubsection{Supersymmetry breaking}

We now observe that the condition of having a cosmological constant {\em now}, together with the concept of the holographic model of 
cosmology that we have developed, naturally imply supersymmetry breaking with a reasonable susy breaking scale. 

Without the cosmological constant, we would have supersymmetry, by construction. Indeed, we have assumed that the the intersecting D6-branes
in $CY_3$ give an MSSM construction, and if our 4 dimensional space were just $\mathbb{R}^4$, then we would have global supersymmetry.  
Moreover, the gravity system was described holographically by an ${\cal N}=1$ 
supersymmetric 3 dimensional field theory, so we should have 4 dimensional ${\cal N}=1$ supergravity, coupled to MSSM. 

But the presence of nonzero energy, the cosmological constant, breaks supersymmetry, and the breaking obviously is in the gravity sector. 
In particular then, that means that the breaking is via a gravitino mass $m_{\psi_\mu}$, and in gravity mediated susy breaking we have generically,
\be
m_{\psi_\mu}\sim \sqrt{G_N} M_S^2\;,
\ee
where $M_S$ is the scale of susy breaking. 

From the point of gravity, the susy breaking is supposed to be nonperturbative in nature, but that means that from the point of view of its 
dual 3 dimensional SYM field theory, it is perturbative in nature, probably something like an FI mechanism or a 
O'Raifeartaugh model. 

But, since the graviton $g_{\mu\nu}$ in cosmology is dual to the field theory energy-momentum tensor $T_{ij}$, its superpartner 
the gravitino $\psi_\mu$ is dual to the supercurrent $Q_{\mu\a}$, and the 
mass of the gravitino is of the order of 
\be
m_{\psi_\mu}\sim \frac{1}{\bar R}\times f(\Delta)\sim \frac{1}{\bar R}\;,
\ee
where $\bar R$ is a scale in the gravity dual,
since the anomalous dimension $\Delta$ corresponding to $Q_{\mu\a}$ is of the order of one (as it is the superpartner of $T_{\mu\nu}$).

We might ask: is the same argument not valid also for the Standard Model (or MSSM) particles?
No,  the SM particles have arbitrary masses since they are not part of a multiplet dual to the adjoint field theory, rather they are added on, a 
small correction (much fewer field modes than the ones dual to gravity) that does not change the gravity dual. 

Next, we need to understand what is the scale $\bar R$, is it the original $R$, $R'$, $a(t)$, $H^{-1}$ or $E_0^{-1}$ from (\ref{LambdaE0}),
since it is a scale in the gravity dual (cosmology, or full 10d metric). It should be a scale defining the geometry from the reheating time until now, 
yet be a constant scale during the same time (so that it doesn't vary: we don't want fundamental scales to vary). Because $E$ 
defines $\Lambda_{\a'}$ in (\ref{lambdaa}) (just before reheating) 
like $E_0$ defines $\Lambda$ in (\ref{LambdaE0}), it is natural to expect $E_0$ to be the scale $1/\bar R$. Moreover, now the geometry 
is certainly defined by $\Lambda$ only, thus by the scale $E_0$.


Since $\Lambda \equiv E_0^4$ gives $E_0\sim 0.1eV$, we obtain
\be
m_{\psi_\mu}\sim E_0 \sim 0.1 eV.
\ee

Note that this does not lead to the "gravitino problem", since we don't mean exactly this value, but rather {\em of the order of} (up to some 
small numerical factors) $E_0$, but in reality its energy density 
is low enough to not contribute enough to the energy density of the Universe ($m_{\psi_\mu}$ within, say, a factor of 
10 or so of $E_0$). The argument only says that the scale appearing in $m_{\psi_\mu}$ is related, up to numerical factors, to the scale 
appearing in $\Lambda$, which is almost the scale $E_0$. 

But then the susy breaking scale is (note that any numerical factors in $m_{\psi_\mu}$ appear under the square root in the equations 
below)
\be
M_S\sim 
\sqrt{10^{-10}GeV\times 10^{19}GeV}\sim 30 TeV\;,
\ee
which is very close to the experimental limit. Indeed, we have argued that the only thing we know for sure (completely model-independent)
is that the susy breaking scale is larger than the maximum energy achieved at particle accelerators.

\subsection{Constraints on extra dimensions and energy scales}

Finally, we should understand the possible values for the size of the extra dimensions now, thus also the values for the KK scale, the string scale
and the 10 dimensional Planck scale. 

We might worry that these sizes vary from the reheating on, but that is not so. As we have argued, from reheating on we 
basically have just 
a usual construction of a RD cosmology, with a truly constant cosmological constant, and MSSM construction for intersecting D6-branes 
in a fixed $CY_3$. 

Before reheating, more precisely during the non-geometric phase that gives CMBR fluctuations, we do have evolving $R(r)$ in the 
10 dimensional metric (\ref{10dmetric}). 

Knowing something about the starting point of the evolution, i.e., the starting point of the 
RG flow of the 3 dimensional field theory, where the $\lambda_{\rm eff}\sim 1$, should correspond to the CMBR fluctuations on horizon scale, 
and also knowing something about the end point of the evolution, reheating, allows us to put constraints on the scale of extra dimensions now. 

The amount of RG flow from the moment when the 3 dimensional YM field theory is strongly coupled (corresponding to the CMBR fluctuations 
at $l\sim 35$), $U_{in}$, to the string energy scale, $U_{\a'}=1/\sqrt{\a'}$, 
\be
\frac{U_{\a'}}{U_{in}}\equiv e^{N_e}\;,
\ee
is a parameter like the exponent of the number of e-folds in inflation (corresponds to this, in the case that the holographic cosmology 
paradigm reduces to inflation), and like that factor, it is restricted by the need to solve the flatness and horizon problems, to be larger than 
$10^{54}$ (see \cite{Nastase:2019rsn,HNSoon}), 
\be
\frac{U_{\a'}}{U_{in}}\geq 10^{54}. \label{UUgeq}
\ee

We have seen in (\ref{Rpr}) that, considering the toy model of $N$ D2-branes (the only one that we can calculate in) to be valid 
from the initial point, where by definition $g^2_{YM}N/U_{in}=1$, to the final point, where $U=U_{\a'}$ (which we can approximate 
as close to the time of reheating), the radius of the compact space (in the 10 dimensional Einstein metric) at reheating (so also 
from reheating until now, since the compact space is then fixed in size, as we explained) is
\be
R'\sim \sqrt{\a'} \left( \frac{g^2_{YM}N}{U_{\a'}}\right)^{-1/16} N^{1/4}\sim \sqrt{\a'} \left(\frac{U_{\a'}}{U_{in}}\right)^{1/16} N^{1/4}\equiv \sqrt{\a'} f.
\ee
From the phenomenological fit to CMBR, $N\simeq 3\times 10^3$, so $N^{1/4}\simeq 10$. Substituting (\ref{UUgeq}), we obtain 
\be
f\geq  10^{4.5}\Rightarrow R'\geq 10^{4.5}\sqrt{\a'}.
\ee

But if we KK compactify on the 6-dimensional space of radius $R'$, we obtain 
\be
M^2_{\rm Pl,4d}=M^8_{\rm Pl,10d}R'^6\Rightarrow 
M_{\rm Pl, 10d}=M^{1/4}_{\rm Pl,4d} R'^{-3/4}.
\ee

Then we get
\be
M_{\rm Pl, 10d}\leq M^{1/4}_{\rm Pl,4d} \times 10^{-3.5}\times \a'^{-3/8}
\ee

The only thing we know for sure in cosmology is what happens from the Big Bang Nucleosynthesis on, so we should impose
\be
\a'^{-1/2}\geq T_{BBN}\simeq 1 MeV. 
\ee

For $1/\sqrt{\a'}=1MeV$, we obtain 
\be
R^{-1}\geq 30 GeV\;,\;\;\; M_{\rm Pl, 10d}\geq 10^5 TeV. 
\ee

On the other hand, it is more reasonable to assume $1/\sqrt{\a'}\geq 10 TeV$, since we didn't see strings at accelerators. 
Assuming $1/\sqrt{\a'}=10 TeV$, we obtain 
\be
R'^{-1}\geq 3\times 10^5 TeV\;,\;\; M_{\rm Pl,10d}\geq 3\times 10^{13}GeV.
\ee

Note that  the string scale $1/\sqrt{\a'}$ and the 10 dimensional Planck scale are related by (\ref{strings}), however, for the 
string coupling we cannot use the D2-brane result (\ref{ND2}), since it is drastically changed in the string scale - reheating region
(if we would use it, we would obtain a nonsensical result, an absurdly low string scale). It would also be drastically changed by 
the inclusion of the fractional D2-branes ($D6_i$ branes on 3-cycles). 

We conclude by noting that, since we mostly used the toy model of $N$ D2-branes only in order to find the constraints on 
$R'$ and $M_{\rm Pl,10d}$, the numbers are not to be taken too seriously. Only the conclusion that large extra dimensions
with TeV strings are consistent with the reheating model, yet smaller ones are better, is.


\section{Conclusions}

In this paper we have filled in the only remaining question about holographic cosmology, namely the analog of reheating, the 
end of the non-geometric phase, and the connection with the usual radiation dominated (RD) cosmology. For reheating, we have 
to use a construction for the Standard Model, in order to find how to couple the geometry to particle physics modes, and we have chosen a 
generic construction of intersecting D6-branes on $CY_3$, that gives the MSSM. Supersymmetry is preserved a priori if we also have 
supergravity cosmology, i.e.  if the 3 dimensional field theory dual is supersymmetric, and we have found that we can modify the best fit to the 
CMBR data to include ${\cal N}=1$ supersymmetric theories. Reheating is introduced by the presence of S-NS5-branes that create the 
intersecting D6-branes, and we have found that it can be done efficiently. 

We have found that we can estimate the reheat temperature, and we can also obtain the
observed cosmological constant $\sim 10^{-123}M_{\rm Pl,4d}^4$ in this model. This $\Lambda$ is a true cosmological constant, and 
with a natural assumption it
implies a breaking of supersymmetry at a scale of about $30 TeV$. From the cosmological constraints on the evolution of the extra dimensional 
space, we find that, in the simple toy model of $N$ D2-branes only, 
for a string scale of $1 MeV$, the KK scale is of $30 GeV$ and the 10 dimensional Planck scale of $10^5 TeV$, whereas in the more reasonable 
case of a string scale of $10 TeV$, the KK scale is $10^5 TeV$, whereas the 10 dimensional Planck scale is about $3\times 10^{13} TeV$. 
That means that large extra dimensions with TeV strings is consistent with the reheating model, but note that smaller ones are preferred. 

There are many things left open for further work. In particular, we have not constructed an explicit intersecting D6-brane model that gives the 
MSSM, and we have not constructed an explicit $D6_i$ on $CY_3$ model that gives the 3 dimensional 
field theory adjoint scalars and fermions needed. This is actually a good thing, since it means there is much flexibility in the result, but it 
would be good to find one concrete construction. In that case, we could also construct explicitly the 3 dimensional field theory fundamental 
modes that correspond to the MSSM ones. 

More importantly, we have used in concrete calculations the toy model of only $N$ D2-branes, without the $D6_i$'s giving adjoint scalars and 
fermions, so one should see if in an explicit model one finds the same results. We have also not described the string theory solution 
(\ref{10dmetric}), we have left it open, with a generic $CY_3$, and a generic $R(r)$, and have not given the other fields. This can be 
seen again as a general picture, that should allow concrete solutions, but we have not provided one.

\section*{Acknowledgements}

I would like to thank Kostas Skenderis for many useful discussions on holographic cosmology. 
I would also like to thank Robert Brandenberger, Rogerio Rosenfeld and Jacob Sonnenschein for 
many relevant comments on the draft, that allowed me to improve it.
My work is supported in part by CNPq grant 301491/2019-4 and FAPESP grants 2019/21281-4 
and 2019/13231-7. I would also like to thank the ICTP-SAIFR for their support through FAPESP grant 2016/01343-7.



\bibliography{holocosmoSM}
\bibliographystyle{utphys}

\end{document}